\documentclass[showpacs,showkeys,11pt,
preprint,preprintnumbers,nofootinbib,
groupedaddress,superscriptaddress,amsmath,amssymb]{revtex4}
\usepackage{amsfonts}
\usepackage{amsmath}
\usepackage{graphics}
\usepackage{epsfig}
\usepackage{url}
\usepackage{multirow}
\usepackage{feynmp}

\newcommand {\be}{\begin{equation}}
\newcommand {\ee}{\end{equation}}
\newcommand {\ba}{\begin{eqnarray}}
\newcommand {\ea}{\end{eqnarray}}

\newcommand {\tanb}{$\tan\beta~$}
\newcommand {\ra}{\rightarrow}
\newcommand {\sinb}{s_{\beta}}
\newcommand {\cosb}{c_{\beta}}
\newcommand {\sina}{s_{\alpha}}
\newcommand {\cosa}{c_{\alpha}}
\newcommand {\sbma}{s_{\beta-\alpha}}
\newcommand {\cbma}{c_{\beta-\alpha}}
\newcommand {\sapb}{s_{\alpha+\beta}}
\newcommand {\capb}{c_{\alpha+\beta}}

\begin{document}
\title{Sources of Charged Higgs Pair Through Double or Triple Higgs Production at Linear Colliders}

\author{Ijaz Ahmed}
\email{Ijaz.ahmed@cern.ch}
\affiliation{National Centre for Particle Physics, University of Malaya, 50603 Kuala Lumpur, Malaysia}
\affiliation{COMSATS Institute of Information Technology (CIIT), Islamabad 44000, Pakistan}


\begin{abstract}
The production of triple Higgs $(H^+H^-H^0)$, $(H^+H^-h^0)$ and pair wise charged Higgs boson $(H^+H^-)$ is studied in the context of future linear colliders within the Two-Higgs Doublet Model (2HDM) type II. The aim is to compare sources of charged Higgs pair through the above processes, i.e., double and triple Higgs production. Cross-sections are calculated at the leading order in 2HDM type II and Minimal Supersymmetric Standard Model (MSSM). Several orders of magnitude ($\sim 10^{4}$) enhancement is observed in 2HDM compared to MSSM, while no sizable enhancement is seen in muon collider versus electron-positron collider. The analysis is based on a heavy charged Higgs with mass above 500 GeV. It is found that double charged Higgs production cross-section (being the same in 2HDM and MSSM) is few femtobarns while the triple Higgs production can not exceed a fraction of femtobarn within the parameter space under study.
\end{abstract}
\maketitle
\section{Introduction}
The main undisputed highlight of Run 1 of the Large Hadron Collider (LHC) at CERN \cite{125atlas,125cms} is the discovery of the Standard Model (SM) Higgs boson. The measured signal strengths are quite in agreement with SM predictions. The mass of the Higgs signal is found to be near 125 GeV, which not only confirms the Higgs mechanism as a right approach towards giving masses to the electroweak particles and gauge bosons, but also puts a question of possibility of existence of further Higgs bosons, as it is still not clear whether the Higgs sector is indeed minimal, containing only a single Higgs doublet. One of the straight forward ways to address such a question is simply to go beyond the SM by adding a second Higgs doublet to the field content of the model. A particularly well-motivated possibility along these lines is the Minimal Supersymmetric Standard Model (MSSM) \cite{MSSM} and the general (unconstrained) Two-Higgs-Doublet Model (2HDM)\cite{2HDM,THDM}. The general 2HDM Higgs sector contains two CP-even neutral Higgs bosons, $h^0$, $H^0$, a CP-odd (pseudo-scaler) neutral Higgs boson, $A^0$, and pair of charged Higgs bosons, $H^\pm$, whereas, h0 is the SM-like Higgs boson and is the candidate for the signal observed at LHC.\\
The purpose of this paper is to take into account all current constraints on the type-II CP-conserving 2HDM parameter space and determine the allowed ranges of the triple and double Higgs couplings to estimate their corresponding cross-sections. This work would prepare the ground for collider studies. Therefore two types of linear colliders, i.e., $e^+e^-$ and $\mu^+\mu^-$ are compared for all processes. Both of these type of processes have been extensively searched during the last years at Tevatron, Large Electron-Positron Collider (LEP) and currently at Large Hadron Collider (LHC). The recent results from LHC exclude a large parameter space of the light charged Higgs, $m_{H^\pm} < 160$ GeV, if BR$(H^{\pm}\ra \tau\nu)=1$ and heavy charged Higgs at tan$\beta>$ 50 \cite{atlasconf2,cmspas}. Furthermore, a remarkable restriction over charged Higgs masses comes from Flavour Changing Neutral Current, (FCNC), radiative B-meson decay, whose branching ratio $BR (b\rightarrow s\gamma)\approx 3\times 10^{-4}$ \cite{Yao} is measured with sufficient precision that becomes sensitive to new physics. The charged Higgs contribution in above branching ratio increases with decreasing $m_{H^{\pm}}$. This channel has been studied by BaBar and Belle collaborations in detail \cite{BaBar,Belle} and the up-to-date limit excludes charged Higgs lighter than 480 GeV at 95\% C.L. \cite{bsgnew}. Therefore the analysis presented throughout the paper is based on $m_{H^{\pm}}\geq 500 GeV$.\\
A phenomenological study of triple Higgs production, including event study and the effect of neutral Higgs masses, $m_H$ and $m_A$, on the cross section can be found in \cite{ijaz} for $m_{H^{\pm}}<500$ GeV. \\
In order to choose right masses for Higgs bosons of 2HDM, one should be aware that the global fit to electroweak measurements requires $\Delta \rho$ to be $O~(10^{-3})$ \cite{rho}. This requirement does not allow large mass splitting between Higgs bosons. Therefore, we adopt degenerate masses for neutral Higgs bosons, i.e., $m_H=m_A=m_{H^{\pm}}$. \\
Regarding the neutral Higgs mass $m_H^0$, recently CMS and ATLAS experiments have excluded a wide range of $m_{H^0}$ in MSSM via a study of $H\rightarrow \tau \tau$ channel at $\sqrt s$ = 8 TeV \cite{cms2,atlas1}. Higgs boson decays to gauge bosons such as $H\ra WW,~ZZ$ and $\gamma\gamma$ have not been studied for MSSM neutral Higgs bosons. They are, however, considered for the light SM Higgs boson as different sources of Higgs boson production. Our chosen MSSM points are outside the LHC excluded area if one sets $m_H=m_A=m_{H^{\pm}}\geq 500$ GeV and \tanb = 10. \\
This study is intended to be suitable at $e^+e^-$ colliders or muon colliders. The muon collider is expected to get the integrated luminosity around 125 $fb^{-1}$ at $\sqrt s$ = 1.5 TeV and 440 $fb^{-1}$ at $\sqrt s$ = 3 TeV. This is a unique machine, to be designed to provide high luminosity, very small energy spread, excellent stability, and good shielding of muon beam decay backgrounds.

\section{Theoretical Framework}
The theoretical basis is a two Higgs doublet model with the general potential as follows \cite{2hdm1,2hdm2}.
\begin{multline*}
\mathcal{V} =  m_{11}^2\Phi_{1}^{\dag}\Phi_{1} + m_{22}^2\Phi_{2}^{\dag}\Phi_{2} - \left[m_{12}^2\Phi_{1}^{\dag}\Phi_{2}+\textnormal{h.c.}\right] \notag \\	
+\frac{1}{2}\lambda_1\left(\Phi_{1}^{\dag}\Phi_{1}\right)^2+\frac{1}{2}\lambda_2\left(\Phi_{2}^{\dag}\Phi_{2}\right)^2 \\
+\lambda_3\left(\Phi_{1}^{\dag}\Phi_{1}\right)\left(\Phi_{2}^{\dag}\Phi_{2}\right) 
+\lambda_4\left(\Phi_{1}^{\dag}\Phi_{2}\right)\left(\Phi_{2}^{\dag}\Phi_{1}\right) \notag \\
+\frac{1}{2}\lambda_5\left(\Phi_{1}^{\dag}\Phi_{2}\right)^2 
+\left[\lambda_6\left(\Phi_{1}^{\dag}\Phi_{1}\right) 
+\lambda_7\left(\Phi_{2}^{\dag}\Phi_{2}\right)\right]\left(\Phi_{1}^{\dag}\Phi_{2}\right)+\textnormal{h.c.}
\end{multline*}

\begin{figure}
 \begin{center}
 \includegraphics[scale=0.8]{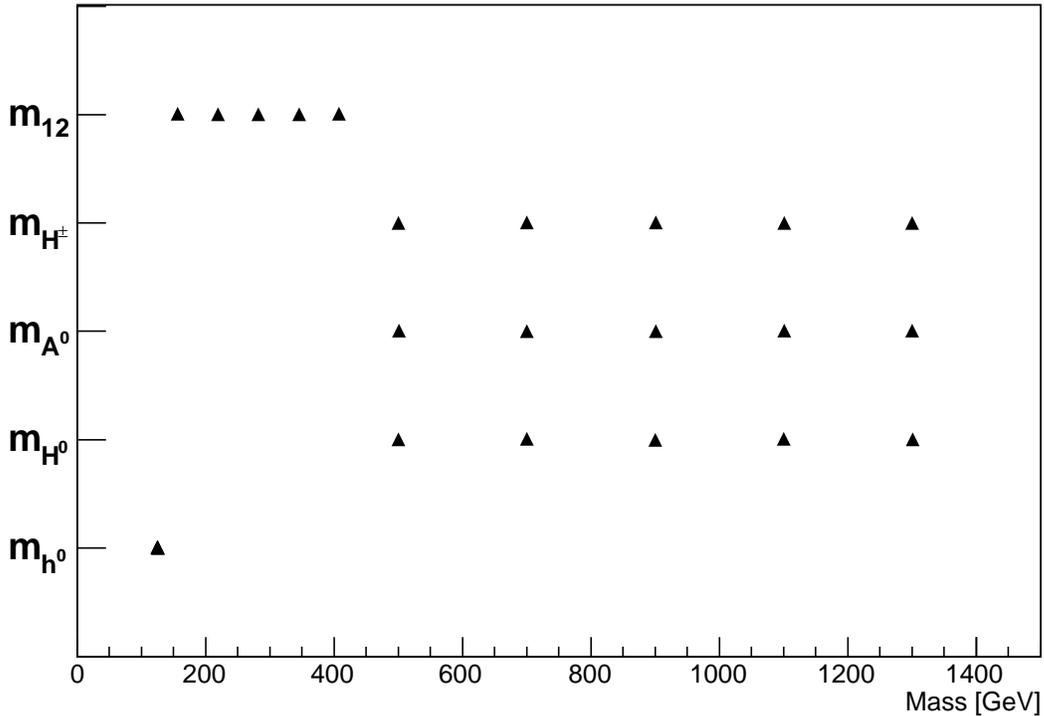}
 \end{center}
 \caption{Higgs boson masses and the value of $m_{12}$ which respect physical requirements on model potential as well as experimental limits and observations on Higgs boson masses.}
 \label{mh}
\end{figure}

\begin{figure}
 \begin{center}
  \includegraphics[scale=0.8]{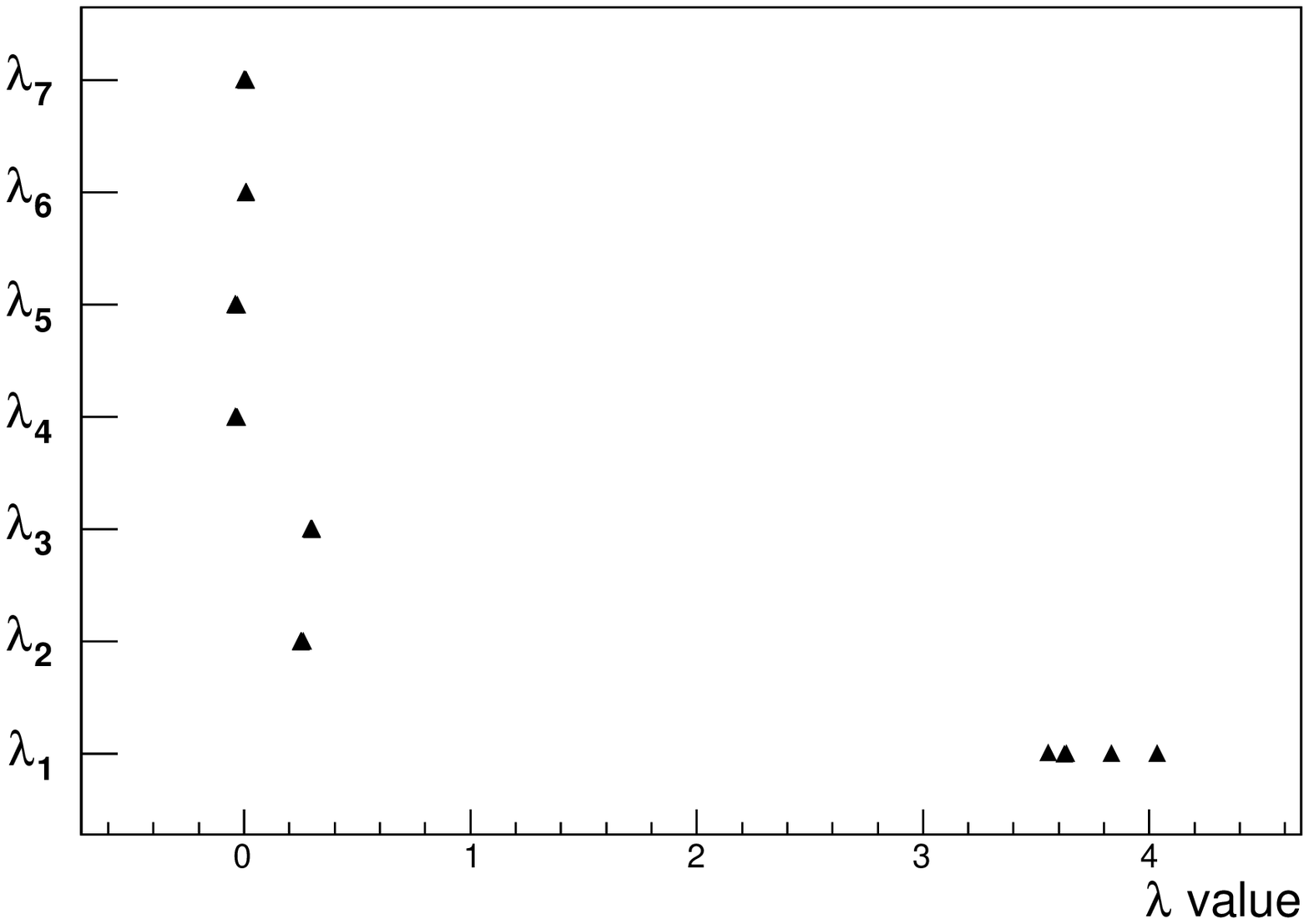}
 \end{center}
 \caption{The values of $\lambda$ parameters which result in Higgs boson masses shown in Fig. \ref{mh}.}
 \label{lambda}
\end{figure}

The free parameters of such a model are  
\be
\lambda_1,\lambda_2,\lambda_3,\lambda_4,\lambda_5,\lambda_6,\lambda_7,m_{12}^{2},\tan\beta 
\ee
in the general basis. The CP violation or Flavor Changing Neutral Currents (FCNC) are not assumed as they are naturally suppressed via the Natural Flavor Conservation (NFC) mechanism by imposing a $Z_{2}$ symmetry on the Lagrangian \cite{weinberg,2hdm3} leads to the 
$\lambda_6=\lambda_7=0$. Furthermore we choose $\sin(\beta-\alpha)=1$ which takes the region of study very close to MSSM parameter space. With the above setting, i.e., \tanb = 10 and $\sin(\beta-\alpha)=1$, free parameters can alternatively be taken as: $m_h,m_H,m_A,m_{H^{\pm}},m_{12}^2,\tan\beta$ because there is a correspondence between Higgs boson masses and $\lambda$ values.
We use 2HDMC package \cite{2hdmc} to ensure that chosen parameters are consistent with current experimental limits and respect also the potential unitarity, perturbativity and stability. A point is chosen if it satisfies all above requirements. Figures \ref{mh} and \ref{lambda} show Higgs boson masses used in the analysis and their corresponding $\lambda$ and $m_{12}$ values extracted from 2HDMC. The choice of the default set of MSSM parameters are $M_{SUSY}$ = 1000 GeV, $\mu$ = 200 GeV, $A_{t}$ = 1000 GeV, $A_{b}$ = 1000 GeV and $A_{\tau}$ = 1000 GeV, which reflect the benchmark $m_{h}^{max}$ scenario.
\section{Charged Higgs Pair Production}
\begin{figure}
 \begin{center}
 \includegraphics[width=0.8\textwidth]{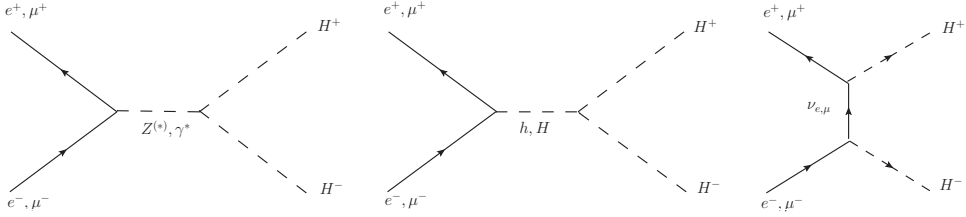}
 \end{center}
 \caption{The possible Feynman diagrams for double Higgs production at lepton colliders.
 \label{HH-diagram}}
 \end{figure}

Double Higgs or pair production of charged Higgs bosons at linear colliders, has been extensively investigated in the literature mainly in MSSM in the context of $e^{+}e^{-}$ and $\mu^{+}\mu^{-}$ colliders, \cite{Djouadi1,Djouadi2,MajidHHeeCollider,MajidHHMuonCollider}. The two types of charged Higgs decay have been adopted in those analyses, i.e., $H^{\pm}\ra \tau\nu$ and $H^{\pm}\ra tb$. One of the week points of this channel is that it is limited by the center of mass energy of the collider and a charged Higgs mass heavier than $\sqrt{s}/2$ is not produced unless a small negligible rate due to off-shell production is considered.
\begin{figure}
 \begin{center}
 \includegraphics[width=0.8\textwidth]{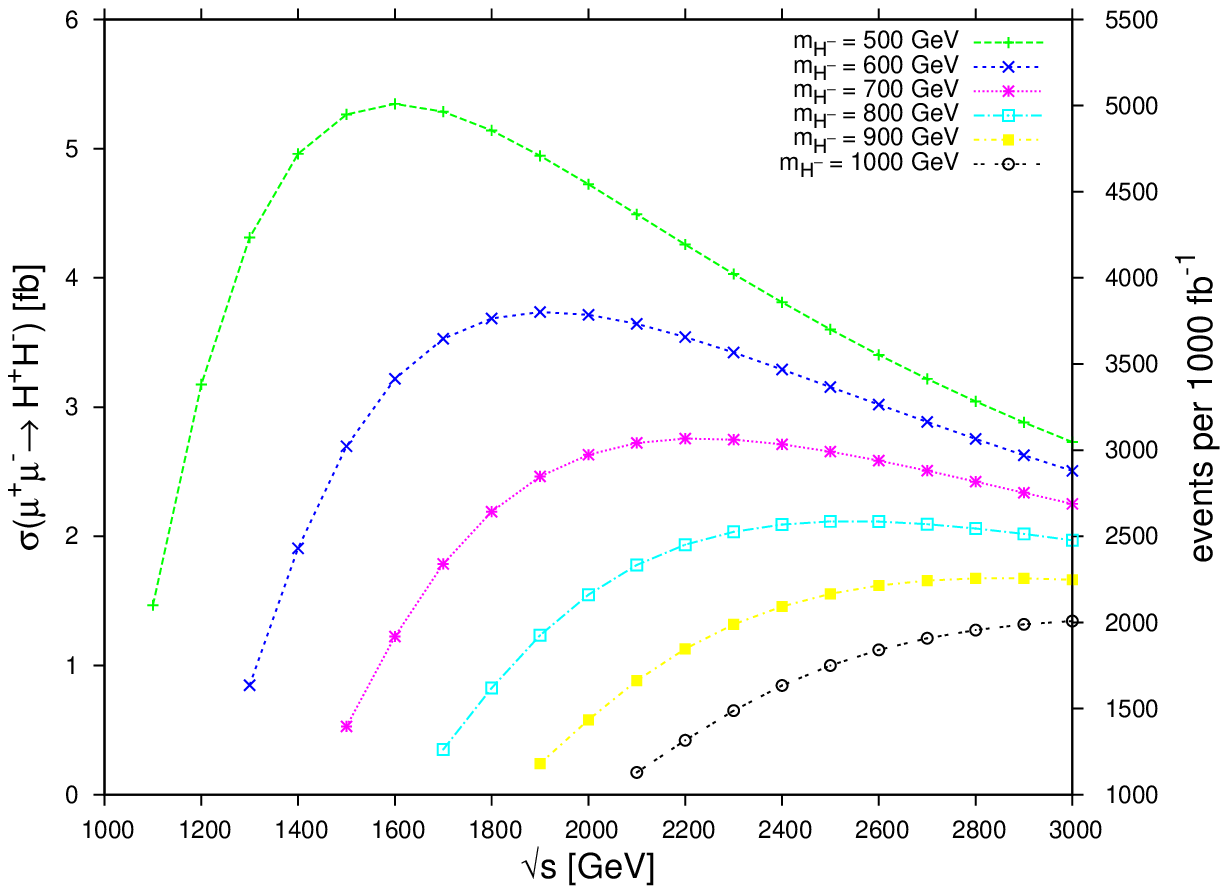}
 \end{center}
 \caption{Total cross-section $\sigma(H^{+}H^{-})$ (in $fb$) as a function of $\sqrt{s}$ for the tree-level Higgs boson pair production at 2HDM or MSSM at a muon collider.}
 \label{HH2HDM}
\end{figure}
In Figure \ref{HH2HDM} the total production cross-section $\sigma (\mu\mu\rightarrow H^{+}H^{-})$ (in pb) is plotted as a function of center-of-mass energy $\sqrt{s}$ (in GeV) for degenerate masses of Higgs bosons in 2HDM type II. The charged Higgs pair production cross section is independent of tan$\beta$ and $m_{A}$ values. As is seen in the Figure \ref{HH2HDM}, the corresponding production rates reach few femtobarns, attaining several thousand events per 1000 $fb^{−1}$ of integrated luminosity. It should be noted that there are Feynman diagrams which could result in enhancement of the production rate. They are $s-$channel diagrams containing neutral Higgs bosons contributions and their couplings with leptons as well as $t-$channel charged Higgs diagrams shown in Fig. \ref{HH-diagram}. However such diagrams have very small contributions to the total cross section and the $Z/\gamma-$mediating diagrams dominate. Since the relevant coupling in this diagram is $ZH^+H^-$ ($e.\cot2\beta$) and $\gamma H^+H^-$ ($e$) and both of them are model independent gauge couplings, the obtained cross sections are the same in 2HDM and MSSM. Table \ref{crosscolmssmDH} compares contribution of different diagrams in the total production cross section at a certain point in parameter space. The conclusion for this observation is that double charged Higgs production ($H^+H^-$) is produced at $\mu^+\mu^-$ and $e^+e^-$ colliders with no sizable difference.  
\begin{table}
\begin{tabular}{|c|c|c|c|c|c|c|}
\hline
& \multicolumn{6}{|c|}{$H^+H^-$}\\
& \multicolumn{3}{|c|}{MSSM} & \multicolumn{3}{|c|}{2HDM}\\
\hline
Collider type & $Z/\gamma$ & $h^{0}/H^{0}$ & Total & $Z/\gamma$ & $h^{0}/H^{0}$ & Total \\
\hline
$\mu^{+}\mu^{-}$ & 2.73 & $1.40\times 10^{-6}$ & 2.73 & 2.74 & $1.41\times 10^{-6}$ &2.74 \\
\hline
$e^{+}e^{-}$ & 2.73 & $1.27\times 10^{-13}$& 2.73 &2.74& $7.61\times 10^{-16}$ &2.74 \\
\hline
\end{tabular}
\caption{Total cross section (in $fb$) of double charged Higgs at $e^+e^-$ and $\mu^+\mu^-$ colliders at $\sqrt s$ = 3 TeV. The Higgs boson masses are $m_H=m_A=m_{H^{\pm}}=500$ GeV. \label{crosscolmssmDH}}
\end{table}
Since production cross section is the same at both 2HDM and MSSM, it is difficult to distinguish the two models using double Higgs ($H^{+}H^{-}$) channel. It should be noted that one can look to single charged Higgs production in association with W boson. This channel has been studied in \cite{a1,a2,myWH} and shows that signal cross section is larger at muon colliders and is almost independent of the charged Higgs mass and increases with \tanb. A similar analysis of this channel at LHC has been repoted in \cite{myWH2}. In the rest of the paper, triple Higgs production is studied. It might be possible that the triple Higgs boson self-interactions play a role in this endeavor, however, as is shown in the next sections, cross section of triple Higgs production ($H^+H^-H^0$, $H^+H^-h^0$) is smaller than double Higgs production ($H^+H^-$). 
\section{Triple Higgs Production}
\begin{figure}
 \begin{center}
 \includegraphics[width=0.8\textwidth]{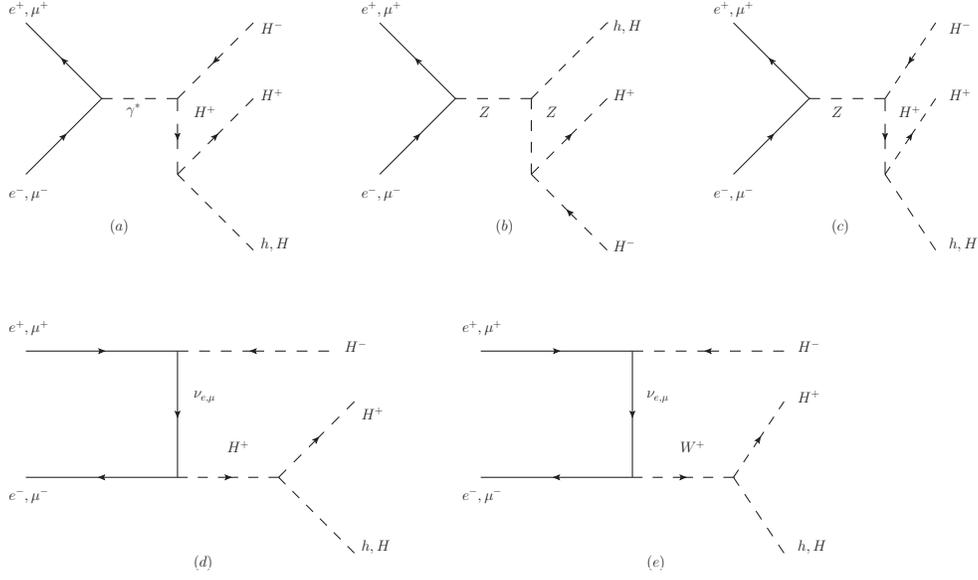}
 \end{center}
 \caption{The possible Feynman diagrams for triple Higgs production at lepton colliders.
 \label{HHH-diagram}}
 \end{figure}

\begin{figure}
 \begin{center}
 \includegraphics[width=0.8\textwidth]{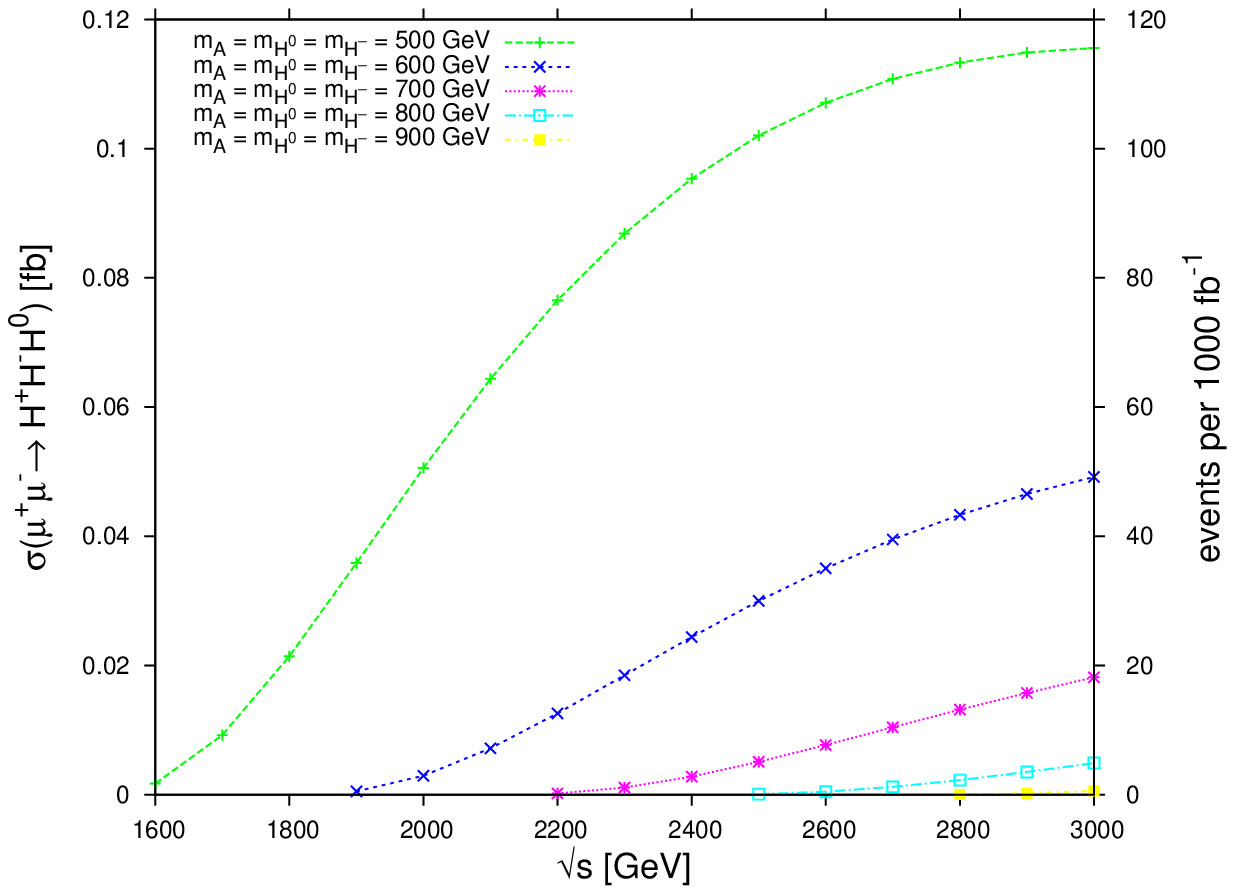} 
 \end{center}
 \caption{Triple Higgs production ($H^+H^-H^0$) within 2HDM type II for various $m_{H^\pm}$ values. Neutral Higgs bosons have masses equal to the mass of the charged Higgs.}
 \label{HHHxsec}
 \end{figure}

\begin{figure}
 \begin{center}
 \includegraphics[width=0.8\textwidth]{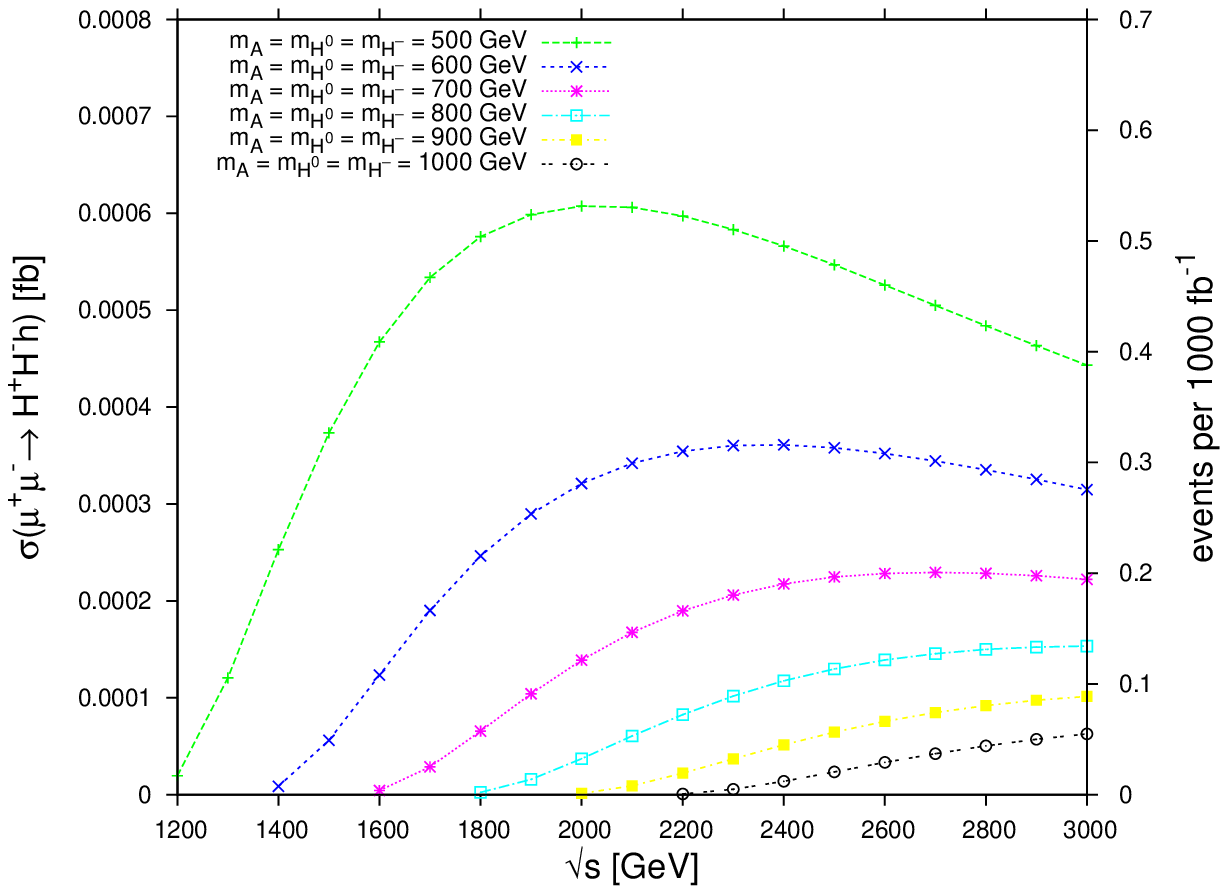}
 \end{center}
 \caption{Triple Higgs production ($H^+H^-h^0$) within 2HDM type II for various $m_{H^\pm}$ values. Neutral Higgs bosons have masses equal to the mass of the charged Higgs.}
 \label{HHhxsec}
 \end{figure}

The triple Higgs production has been studied in different papers in the context of linear colliders \cite{Kanemura,Arhrib,Ferrera3,Lopez,Guti1}. The production processes under consideration are $H^{+}H^{-}H^{0}$ and $H^{+}H^{-}h^{0}$ in $\mu^{+}\mu^{-}$ annihilations within the 2HDM type II with their corresponding trilinear Higgs bosons couplings given in the Eqns. \ref{HHH} and \ref{HHh} together with their corresponding MSSM values in Eqns. \ref{mssmcoupling1} and \ref{mssmcoupling2}, where usual abreviations such as $s_W=\sin\theta_{W}$ and $s_{\beta}=\sin\beta$ have been used \cite{dubinin}. \\
\begin{widetext}
\begin{equation}
H^{\pm}H^{\pm}H^0 (2HDM):~\frac{-e}{m_W.s_W.s^2_{2\beta}}(\cosb^3s_{2\beta}\sina m_{H}^2+\cosa s_{2\beta}\sinb^3m_{H}^2-2\sapb\mu_{12}^2+\cbma s_{2\beta}^2m_{H^{\pm}}^2) 
\label{HHH}
\end{equation}
\end{widetext}
\begin{widetext}
\begin{equation}
H^{\pm}H^{\pm}h^0 (2HDM):~\frac{e}{m_W.s_W.s^2_{2\beta}}(2\capb\mu_{12}^2-\cosa \cosb^3 s_{2\beta}m_{h}^2+s_{2\beta}\sina s_{\beta}^3m_{h}^2-s_{2\beta}^2\sbma m_{H^{\pm}}^2) 
\label{HHh}
\end{equation} 
\end{widetext}
\begin{widetext}
\begin{equation}
H^{\pm}H^{\pm}H^0(MSSM)=\frac{-e.m^2_Z}{2m_W s_W }(c_{2W}\cbma+s_{2\beta}\sapb)
\label{mssmcoupling1}
\end{equation}
\end{widetext}
\begin{widetext}
\begin{equation}
H^{\pm}H^{\pm}h^0(MSSM)=\frac{-e.m^2_Z}{2m_W s_W }(2 c^2_W\sbma+c_{2\beta}\sapb)
\label{mssmcoupling2}
\end{equation}
\end{widetext}
Figure \ref{HHH-diagram} shows Feynman diagrams of triple Higgs production. In this case there is \tanb dependence, contrary to the case of $H^+H^-$ production. The $H^+H^-H^0$ coupling effectively grows as $~$tan$\beta$ or $~$cot$\beta$ for tan$\beta >>$ 1 or tan$\beta << $1 respectively. The corresponding cross-section can vary either by tan$^{2}\beta$ at larger tan$\beta$ values or by cot$^{2}\beta$ at small tan$\beta$  values respectively in contrast to the situation in MSSM, where the triple Higgs coupling undergoes radiative corrections \cite{Hollik} and does not have any possible source of enhancement as can be seen from Eqn.\ref{mssmcoupling1} showing tree level coupling in MSSM. The Eqn. \ref{mssmcoupling1} clearly indicate that the couplings are naturally gauge like and hence the expected cross-section remains rather small \cite{Djouadi2}. Figures \ref{HHHxsec} and \ref{HHhxsec} show cross sections of $H^+H^-H^0$ and $H^+H^-h^0$ production for different set of charged Higgs masses and center of mass energies.\\
One can see the contribution of different parts of the set of diagrams and compare them in a table as was done for double Higgs production. Table \ref{crosscolmssm} shows such information for $H^+H^-H^0$ and $H^+H^-h^0$ processes, where a comprehensive comparison between 2HDM and MSSM results, for $\mu^+\mu^-$ and $e^+e^-$ colliders has been done.\\
At this stage, a comparison between the two types of processes, i.e., double and triple Higgs production, is performed by finding a center of mass energy which gives the maximum allowed cross-section. In Table \ref{crossECM}, the $\sigma_{max}~(H^+H^-H^0)$, $\sigma_{max}~(H^+H^-h^0)$ and $\sigma(H^+H^-)$ at two different center of mass energies ($\sqrt s$ = 1 TeV and $\sqrt s$ = 1.5 TeV) are listed. The abreviated name "NP" stands for "Not Possible", due to not enough center of mass energy to produce a given event.
As is seen from Table \ref{crossECM}, the MSSM cross-sections for both triple Higgs cases are extremely small reaching the largest value of $\sigma_{max}~(H^+H^-H^0)~\approx~10^{-6}$ $fb$ and $\sigma_{max}~(H^+H^-h^0)~\approx~10^{-5}$ $fb$. Comparing 2HDM and MSSM cross sections, it is observed that one could reach a factor of $10^{5}$ enhancement in 2HDM cross section of $H^+H^-H^0$, while for $H^+H^-h^0$, the 2HDM cross section is typically $10$ times larger than the corresponding MSSM value. 
On the basis of these results, the triple Higgs boson channels are generally much more promising in 2HDM than MSSM framework, but they do not reach the level of $H^+H^-$ cross section. The double Higgs production cross-section is substantially larger than triple Higgs channels either in MSSM or 2HDM. Therefore any study of the charged Higgs at linear colliders should be based on double Higgs production, unless a large integrated luminosity and center of mass energy is available for production of triple Higgs processes. Such events can only be considered as complementary processes to shade some light on triple Higgs couplings.
\begin{figure}
 \begin{center}
 \includegraphics[width=0.8\textwidth]{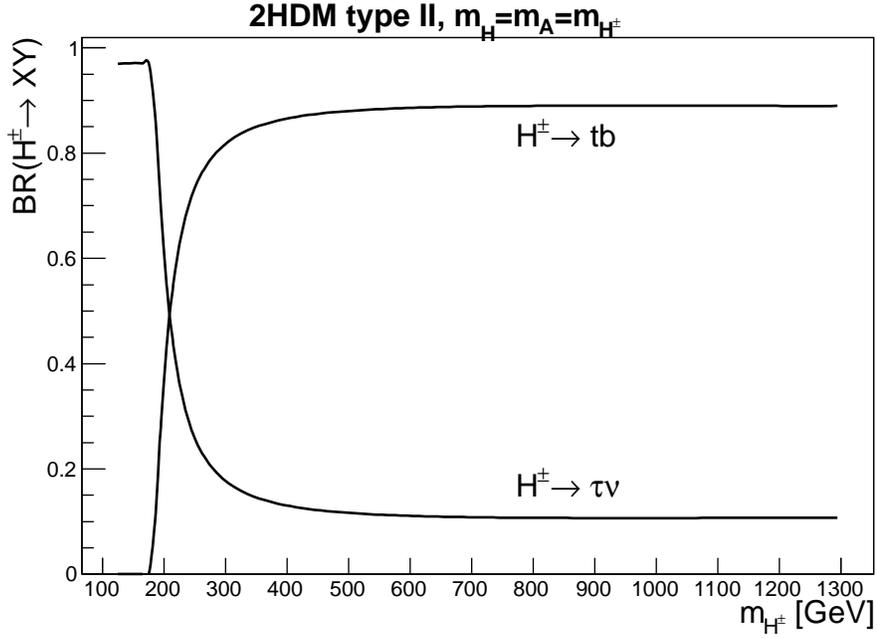}
 \end{center}
 \caption{Branching ratio of charged Higgs decays in 2HDM type II. 
 \label{BR}}
 \end{figure} 
\begin{table}
\begin{tabular}{|c|c|c|c|c|c|c|c|}
\hline
& & \multicolumn{3}{|c|}{$H^+H^-H^0$} & \multicolumn{3}{|c|}{$H^+H^-h^0$}\\
\hline
Collider type & Model type &$Z/\gamma$ & $h^{0}/H^{0}$ & Total & $Z/\gamma$ & $h^{0}/H^{0}$ & Total \\
\hline  
$\mu^{+}\mu^{-}$ & MSSM &$1.93\times 10^{-7}$ & $3.49\times 10^{-6}$& $3.69\times 10^{-6}$ &$1.69\times 10^{-5}$ &$2.85\times 10^{-8}$ & $1.69\times 10^{-5}$ \\
$\mu^{+}\mu^{-}$ & 2HDM & $0.116$ & $2.24\times 10^{-6}$ & $0.116$ & $4.44\times 10^{-4}$ &$1.90\times 10^{-10}$ & $4.44\times 10^{-4}$  \\
\hline
$e^{+}e^{-}$ & MSSM &$1.93\times 10^{-7}$ & $7.77\times 10^{-11}$ & $1.93\times 10^{-7}$ & $1.69\times 10^{-5}$ &$6.34\times 10^{-13}$ & $1.69\times 10^{-5}$ \\
$e^{+}e^{-}$ &2HDM &$0.116 $& $5.10\times 10^{-11}$ & $0.116$ &$4.44\times 10^{-4}$ & $1.03\times 10^{-19}$ & $4.44\times 10^{-4}$ \\
\hline
\end{tabular}
\caption{Total cross-section (in $fb$) comparison between electron-positron and muon-muon colliders as well as among MSSM and 2HDM at $\sqrt s$ = 1.5 TeV. \label{crosscolmssm}}
\end{table}
\begin{table}
\begin{tabular}{|c|c|c|c|c|c|}
\hline
\multicolumn{6}{|c|}{}\\
Process & $m_{H^{\pm}}$ & $\sigma_{max}$ (1.5 TeV) & $\sigma_{max}$ (3 TeV) & $\sigma_{max}$ (1.5 TeV) & $\sigma_{max}$ (3 TeV) \\
& GeV & MSSM & MSSM & 2HDM & 2HDM  \\
\hline
$\mu^{+}\mu^{-}\rightarrow H^{+}H^{-}$ & 500 & 5.26 & 2.72 & 5.26 & 2.72 \\
$\mu^{+}\mu^{-}\rightarrow H^{+}H^{-}H^{0}$ & 500 & NP & $3.68\times 10^{-6}$ & NP & 0.12\\
$\mu^{+}\mu^{-}\rightarrow H^{+}H^{-}h^{0}$ & 500 & $1.44\times 10^{-5}$ & $1.69\times 10^{-5}$ & $3.73\times 10^{-4}$ & $4.44\times 10^{-4}$ \\  
\hline  
$\mu^{+}\mu^{-}\rightarrow H^{+}H^{-}$ & 700 & 5.28 & 2.25 & 5.28 & 2.25 \\
$\mu^{+}\mu^{-}\rightarrow H^{+}H^{-}H^{0}$ & 700 & NP & $3.86\times 10^{-7}$ & NP & $1.81\times 10^{-2}$\\
$\mu^{+}\mu^{-}\rightarrow H^{+}H^{-}h^{0}$ & 700 & NP & $1.13\times 10^{-5}$ & NP & $2.2\times 10^{-4}$ \\
\hline
$\mu^{+}\mu^{-}\rightarrow H^{+}H^{-}$ & 900 & NP & 1.66  & NP & 1.66\\
$\mu^{+}\mu^{-}\rightarrow H^{+}H^{-}H^{0}$ & 900 & NP & $3.86\times 10^{-8}$ & NP  & $5.77\times 10^{-4}$ \\
$\mu^{+}\mu^{-}\rightarrow H^{+}H^{-}h^{0}$ & 900 & NP & $5.27\times 10^{-6}$ & NP & $1.02\times 10^{-4}$\\
\hline
\end{tabular}
\caption{The maximum cross-section (in $fb$) for the leading order double Higgs and triple Higgs processes within MSSM and 2HDM at two different center of mass energies $\sqrt s$ = 1.5 TeV and $\sqrt s$ = 3 TeV.\label{crossECM}}
\end{table}
\section{Cross section at $e^+e^-$ and $\mu^+\mu^-$ Colliders}
In this section, cross sections of three selected channels at two possible choices of colliders, i.e., $e^+e^-$ and $\mu^+\mu^-$ are compared. As shown in Figure \ref{HH-diagram}, the charged Higgs pair production may proceed through three types of diagrams. The left diagram includes s-channel electroweak propagators while the middle one consists of neutral Higgs bosons as propagator. The CP odd neutral Higgs boson does not contribute in the production cross section due to having zero coupling with charged Higgs pair. Therefore we only take into account the electroweak and CP-even neutral Higgs bosons contribution. The amount of cross section coming from each diagram may be considerably important probably after a long run of any one of these two colliders in future. There is also a t-channel process (the right diagram in Fig. \ref{HH-diagram}), which has negligibly small contribution at both $e^+e^−$ and $\mu^+\mu^−$ colliders due to the small Yukawa coupling between the charged Higgs and the leptons. The Table \ref{crosscolmssmDH} demonstrates that $H^+H^-$ cross section is dominated by the $Z/\gamma$ mediating diagram (the left diagram in Fig. \ref{HH-diagram}) and no difference is obtained between $e^+e^-$ and $\mu^+\mu^-$ even though by introducing the non zero masses of electron and muon and their corresponding couplings in the Lagrangian. There is in fact a large enhancement in $h/H$ mediating diagram, however, the total cross section is not affected by this enhancement due to the negligible contribution of collider dependent diagrams.\\ 
The same conclusion holds for the triple Higgs production processes shown in Figure \ref{HHH-diagram}. The diagram with only $Z/\gamma$ propagator (Fig. \ref{HHH-diagram}-b) has the dominant contribution and as seen in Table \ref{crosscolmssm} the effect of collider dependent diagrams, i.e., the ones with neutral or charged Higgs coupling with leptons, is small.\\ 
In summarizing the results we can conclude that for the charged Higgs searches the most suitable channel from cross section point of view is charged Higgs pair production having largest cross-section of the order of few $fb$ in MSSM and 2HDM, while the selected triple Higgs processes ($H^+H^-H^0$, $H^+H^-h^0$) have very tiny cross-sections. The triple Higgs searches could benefit from the neutral Higgs decay to $b\bar{b}$ which can be identified by $b-$taging algorithm. However, due to the larger particle multiplicity compared to $H^+H^-$ production, selection efficiencies of the triple Higgs are expected to be small. The final state to analyze for both double and triple Higgs cases could be defined by selecting one of the two possible decay channels of the charged Higgs, i.e., $H^{\pm}\ra\tau\nu$ and $H^{\pm}\ra tb$. As Figure \ref{BR} shows, the charged Higgs branching ratio of decay to $\tau\nu$ goes down to the level of 0.1, for $m_{H^{\pm}}\geq 500$ GeV, while decay to $tb$ receives a branching ratio of 0.9. Since cross sections are small, especially, for triple Higgs production, the main channel to study is expected to be the one with top and bottom quarks in the hard scattering final state. Of course in this case a large number of particles is produced due to the decay of the top quark to W boson and its subsequent decay to leptons or light jets. One the contrary, the final state with $\tau$ leptons involved, may be identified using a $\tau$ identification algorithm but it certainly suffers from the very tiny number of events in hand. In any case the double Higgs production is expected to be more promising than the triple Higgs production due to the larger cross section. 
\section{Summary}
The most promising channels were studied for the charged Higgs production through double or triple Higgs production at future $e^+e^-$ and $\mu^+\mu^-$ linear colliders in the framework of CP-conserving Two Higgs Doublet Model type II and Minimal Supersymmetric Standard Model. The sizes of double and triple Higgs production cross-sections were computed both in MSSM and 2HDM by respecting all the current experimental and phenomenological constraints on Higgs masses, requirement of perturbativity and vacuum stability condition. 
The maximum cross-sections at two different center of mass energies of 1.5 TeV and 3 TeV were obtained to compare these processes between MSSM and 2HDM where several order of magnitude difference was observed. A comparison between $e^{+}e^{-}$ and $\mu^{+}\mu^{-}$ colliders was also performed leading to no sizable difference at both colliders. The conclusion is that double Higgs production cross section is much larger than triple Higgs, thus keeping it as the main sourse of the charged Higgs at linear colliders. There are diagrams which depend on the collider type (incoming lepton) but they do not have a sizable contribution to affect the total cross section. This fact is observed for both double and triple Higgs productions. The double Higgs production is less limited by the center of mass energy of the system compared to the triple Higgs production and thus can probe heavier charged Higgs bosons at a given center of mass energy. The final state to choose for a data analysis would be $H^+H^- \ra t\bar{b}\bar{t}b$ and $H^+H^-H^0(H^+H^-h^0) \ra t\bar{b}\bar{t}bb\bar{b}$ for the top and bottom final states, and, $H^+H^- \ra \tau^+\nu\tau\bar{\nu}$ and $H^+H^-H^0(H^+H^-h^0) \ra \tau^+\nu\tau\bar{\nu}b\bar{b}$ for $\tau$ and missing transverse energy final states. While final states with $\tau$ and missing transverse energy are suitable for light charged Higgs searches, a heavy charged Higgs with $m_{H^{\pm}}\geq500$ GeV would require a final state including top and bottom quarks.

\end{document}